%% file: paper_eprint.tex
\documentclass[twocolumn,showpacs,aps,prl,superscriptaddress]{revtex4}

\usepackage{graphicx}
\usepackage{dcolumn}
\usepackage{amsmath}
\usepackage{epsfig}

\input pubboard/babarsym
\input pubboard/figures
\newcommand{\BABARPubYear}    {03}
\newcommand{\BABARPubNumber}  {39}

\newcommand{\SLACPubNumber} {10296}

\def\Journal#1#2#3#4{{#1} {\bf #2}, #3 (#4)}

\def\NIMA{{\em Nucl. Instrum. Methods} A}

\def\PLB{{\em Phys. Lett.}  B}
\def\PRL{\em Phys. Rev. Lett.}
\def\PRD{{\em Phys. Rev.} D}

\def\bmunu      {\ensuremath{\Bp \to \mup \num}\xspace}

\def\blnu       {\ensuremath{\Bp \to \ellp \nul}\xspace}

\def\figurebox#1#2#3{%
    \def\arg{#3}%
    \ifx\arg\empty
    {\hfill\vbox{\hsize#2\hrule\hbox to #2{\vrule\hfill\vbox to #1{\hsize#2\vfill}\vrule}\hrule}\hfill}%
    \else
    {\hfill\epsfbox{#3}\hfill}%
    \fi}

\long\def\inst#1{\par\nobreak\kern 4pt\nobreak
    {\it #1}\par\vskip 10pt plus 3pt minus 3pt}

\begin{document}

\preprint{\babar-PUB-\BABARPubYear/\BABARPubNumber} 
\preprint{SLAC-PUB-\SLACPubNumber} 

\begin{flushleft}
  \babar-PUB-\BABARPubYear/\BABARPubNumber \\
  SLAC-PUB-\SLACPubNumber\\[10mm]
\end{flushleft}

\title{
{\large \bf
Search for the Rare Leptonic Decay \boldmath{\bmunu} }
}

\input pubboard/authors_oct2003

\date{\today}

\begin{abstract}
We have performed a search for the rare leptonic decay \bmunu with data 
collected at the \FourS resonance by the \babar\ experiment at the \pep2\ storage ring.
In a sample of 88.4 million \BB pairs, we find
no significant evidence for a signal and set an upper limit on the branching 
fraction $\BR(\bmunu) < 6.6\times10^{-6}$ at the 90\% confidence level. 
\end{abstract}


\maketitle


\setcounter{footnote}{0}


The study of the purely leptonic decays \blnu ( $\ell$ = \electron, \mmu, or \mtau ) can provide 
sensitivity to poorly 
constrained Standard Model (SM) parameters and also act as a probe of new physics. In the SM, 
these decays proceed through \W-boson annihilation with a branching fraction given by 
\begin{eqnarray}
 \BR(\blnu) = \frac{G_{F}^{2} m_{B} m_{\ell}^{2}} {8\pi} 
 \biggl( 1- \frac{m_{\ell}^{2}}{m_{\B}^{2}} \biggr)^{2} \fsubb^{2} \Vub^{2} 
 \tau_{\Bp}, \nonumber
\end{eqnarray}
where $G_F$ is the Fermi coupling constant, $m_{\ell}$ and $m_B$ are 
the lepton and \B meson masses, and $\tau_{\Bp}$ is the \Bp lifetime.
The decay rate is sensitive to the product of the Cabibbo-Kobayashi-Maskawa (CKM) 
matrix element \Vub and the \B decay constant \fsubb, 
which is proportional to the wave function for zero separation between the quarks.
Currently, our best understanding of \fsubb comes from lattice gauge calculations where the theoretical 
uncertainty is roughly 15\%~\cite{Ryan:2001ej}.
This uncertainty is a significant limitation on the extraction of \Vtd from precision
\BzBzb mixing measurements~\cite{Battaglia:2003in}. 
Observation of \blnu could provide the first direct measurement of \fsubb.

In this note, we present a search for the decay \bmunu (charge conjugation is implied throughout this paper). 
This decay is highly suppressed due to the dependence on \Vub\!$^2$ and $m_{\ell}^2$ (helicity suppression). 
Assuming \Vub = 0.0036~\cite{PDG} and \fsubb = 198 \mev~\cite{Ryan:2001ej}, the SM prediction for the \bmunu 
branching fraction is roughly $4\times10^{-7}$. The current best published limit, from the CLEO collaboration, is   
$\BR(\bmunu) < 2.1\times10^{-5}$ at the 90\% confidence level~\cite{Cleo}. 

The \blnu decay modes are also potentially sensitive to physics beyond the SM. 
For example, in two-Higgs-doublet models such as the Minimal Supersymmetric Standard Model (MSSM), these decays can proceed at tree-level 
via an intermediate $H^{\pm}$, providing a possible enhancement up to current experimental limits~\cite{Hou}.  
Similarly, in $R$-parity violating extensions of the MSSM, \blnu may be mediated by scalar 
supersymmetric particles~\cite{Baek:1999ch,Akeroyd:2002cs}. Hence, upper limits 
on the \blnu branching fractions constrain the $R$-parity violating couplings. 


The data used in this analysis were collected with the \babar\ detector
at the \pep2 storage ring. 
The data sample consists of an integrated luminosity
of 81.4 \invfb accumulated at the \FourS resonance (``on-resonance'') and 9.6 \invfb accumulated 
at a center-of-mass (CM) energy about 40 \mev below the \FourS resonance (``off-resonance'').
The on-resonance sample corresponds to 88.4 million \BB pairs. 

The \babar\ detector is optimized for the asymmetric collisions at \pep2 and 
is described in detail elsewhere~\cite{babar}. Charged particle trajectories
 are measured with a 5-layer double-sided silicon vertex tracker (SVT) and
a 40-layer drift chamber (DCH), which are contained in the 1.5 T magnetic field of a 
superconducting solenoid. 
A detector of internally reflected Cherenkov radiation provides identification of charged
kaons and pions. 
The energies of neutral particles are measured by an electromagnetic calorimeter (EMC) 
consisting of 6580 CsI(Tl) crystals. 
The flux return of the solenoid is instrumented with resistive plate chambers 
to provide muon identification (IFR). 
A Monte Carlo (MC) simulation of the \babar\ detector based on {\tt GEANT}4~\cite{geant4}
was used to optimize the signal selection criteria and evaluate the signal efficiency.


The \bmunu decay produces a mono-energetic muon in the \B rest frame 
with $p_{\mu} \approx m_B/2$. Since the neutrino goes undetected, we assume that all remaining particles are associated 
with the decay of the other \B in the event, which we denote the ``companion'' \B. 
Signal events are selected using the kinematic variables \DeltaE $ = E_{B}^*-E_{\rm b}^*$ 
and energy-substituted mass, \mes $= \sqrt{E_{\rm b}^{*2}-{\bf p}_B^{*2}}$
where ${\bf p}_B^*$ ($E_B^*$) is the momentum (energy) of the reconstructed 
companion \B and $E_{\rm b}^*$ is the beam energy, all in the \FourS rest frame. 
We require \mes to be consistent with the \B meson mass, and the energy of the 
companion \B to be consistent with $E_{\rm b}^*$ resulting in \DeltaE $\simeq$ 0. 

To reduce non-hadronic backgrounds, we select events that contain at least four charged tracks and have a 
normalized second Fox-Wolfram moment~\cite{FoxWolf} less than 0.98. 
Muon candidates are required to penetrate at least 2.2 interaction lengths of material in the IFR, 
have a measured penetration within 0.8 interaction lengths of that expected for a muon, and have an associated energy in the EMC 
consistent with that of a minimum-ionizing particle.
The muon track must have at least 12 DCH hits, momentum transverse to the beam axis $p_{\perp} > 0.1$ \gevc, 
and a point of closest approach to the interaction point that is within 10\cm along the beam axis and less than 1.5\cm in the transverse plane.
For each muon candidate with momentum between 2.25 and 2.95 \gevc in the CM frame, we attempt to reconstruct the companion \B.
 
The companion \B is formed from all charged tracks satisfying the above criteria regarding 
the distance of closest approach to the interaction point. 
It also includes all neutral calorimeter clusters with energy greater than 30 \mev. 
Particle identification is applied to the charged tracks 
to identify electrons, muons, kaons and protons and the resulting mass hypotheses are applied to these tracks to improve the 
\DeltaE resolution. Events with additional identified leptons from the companion \B are discarded since  
they typically arise from semi-leptonic \B or charm decays and indicate the presence of additional neutrinos.
  
Once the companion \B is reconstructed, we calculate the muon momentum in the rest frame of the signal \B. 
We assume the signal \B travels in the direction opposite that of the companion \B momentum in the \FourS rest frame 
with a momentum determined by the two-body decay \FourS\to\BpBm.
For signal muons, the $p_{\mu}$ distribution peaks at 2.64 \gevc with an RMS of about 100 \mevc.

Backgrounds may arise from any process producing charged tracks in the 
momentum range of the signal muon.
The two most significant backgrounds are \B semi-leptonic decays involving
$b \rightarrow u \mu {\bar \nu}$ transitions where the endpoint of the muon 
spectrum approaches that of the signal, and non-resonant $q\bar{q}$ (``continuum'') events 
where a charged pion is mistakenly identified as a muon. The pion misidentification rate has been studied using a 
pion control sample taken from \epem\to\tautau events in the data. In the momentum and polar angle region relevant for \bmunu, 
the misidentification probability is estimated to be 2\%.
In order for continuum events to populate the signal region of \DeltaE and \mes, there must be 
significant missing energy due to particles outside the detector acceptance, unreconstructed neutral hadrons, or 
additional neutrinos. 
The muon momentum spectrum of the background decreases
with increasing momentum so we apply an asymmetric cut about the signal peak, $2.58 < p_{\mu} < 2.78$ \gevc.  

The continuum background is further suppressed using event-shape variables. These events
tend to produce a jet-like event topology whereas \BB events tend to be spherical. 
We define a variable, $\theta_T^*$, which is the angle between the 
muon candidate momentum and the thrust axis of the companion \B in the CM frame. For continuum background,
$|\cos\theta_T^*|$ peaks sharply near one while the distribution is nearly flat for signal decays. 
By requiring $|\cos\theta_T^*| < 0.55$,
we remove approximately 98\% of the continuum background while retaining 54\% of the signal decays.
We also use the polar angle of the missing momentum vector in the laboratory frame, $\theta_{\nu}$,
to discriminate against continuum backgrounds.
In continuum events, 
the missing momentum is often due to particles that were outside the detector acceptance. 
Therefore, we require $|\cos\theta_{\nu}| <$ 0.88 so that 
the missing momentum is directed into the detector's fiducial volume. 
  
We select \bmunu signal candidates with simultaneous requirements on \DeltaE and \mes, thus 
forming a ``signal box'' defined by $-$0.75 $<$ \DeltaE $<$ 0.5 \gev and \mes $>$ 5.27 \gevcc. The dimensions of the signal box, as well as the 
above requirements on $p_{\mu}$, $|\cos\theta_T^*|$ and $|\cos\theta_{\nu}|$, were determined using an optimization 
procedure that finds the combination of cuts that maximizes the quantity 
$S/\sqrt{S+B}$ where $S$ and \B are the signal and background yields in the MC simulation respectively. 
The signal branching fraction was set to the SM expectation during the optimization procedure. 
In the MC simulation, 24.5\% of signal decays 
passing all previous cuts fall within the signal box. After applying all selection criteria, the 
\bmunu efficiency is determined from the simulation to be (2.24$\pm$0.07)\%, where the uncertainty is due to MC statistics. 

In addition to the signal box, we have defined a slightly larger blinding box and three sideband regions. The boundaries of these regions 
in the (\DeltaE, \mes) plane are listed in Table~\ref{tab:boxes}. The data within the blinding 
box were kept hidden until the analysis was completed in order to avoid the introduction of bias in the event-selection process. 

\begin{table}
 \caption{  \label{tab:boxes}The boundaries of the signal box and various sidebands defined for this analysis. }
 \begin{ruledtabular}
 \begin{tabular}{lcc}  
   region                       & \DeltaE (\gev)            & \mes (\gevcc)  \\
\hline
   signal box                   & [ -0.75, 0.50 ]             & $>$ 5.27 \\
   blinding box                 & [ -1.30, 0.70 ]             & $>$ 5.24 \\  
   fit sideband                 & [ -0.75, 0.50 ]             & [ 5.10, 5.24 ] \\
   \DeltaE sideband (bottom) & [ -3.00, -1.30 ]            & $>$ 5.10 \\
   \DeltaE sideband (top)    & [ 0.70, 1.50 ]              & $>$ 5.10 \\
 \end{tabular}
 \end{ruledtabular}
\end{table}

\begin{figure}
  \includegraphics[bb=20 30 460 460, height=6cm]{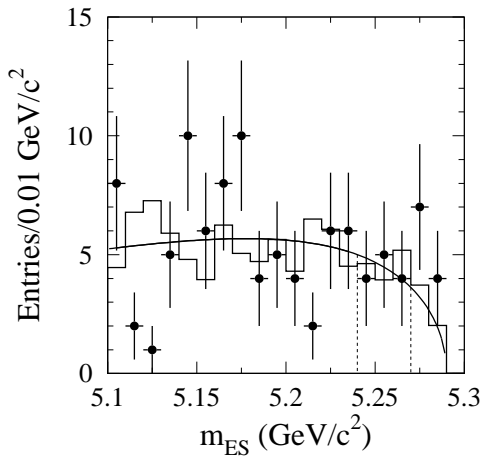}
  \caption{\label{fig:fit}Results of the ARGUS fit to the on-resonance data satisfying  
                          $-0.75 < \DeltaE < 0.5$\gev. The two dashed lines indicate the lower boundaries
                          of the blinded region and signal box at 5.24 \gevcc and 5.27 \gevcc, respectively.
                          The fit is performed only on the region 5.10 $<$ \mes $<$ 5.24 \gevcc and extrapolated 
                          into the signal region to estimate the background. The histogram shows the sum of all
                          simulated background sources normalized to the on-resonance luminosity.}
\end{figure}

We estimate the background in the signal box assuming that the \mes distribution is described by the ARGUS function~\cite{Argus}. 
This assumption is consistent with the observed distributions in the MC simulation as well as the data in the \DeltaE sidebands. 
The single parameter of the ARGUS function ($\zeta$) is determined from an unbinned maximum likelihood fit using the data in the 
fit sideband defined in Table~\ref{tab:boxes}. The ARGUS shape ($A$) is extrapolated 
through the signal box and constrained to be zero at the endpoint, which is fixed at $E_{\rm b}^*=5.29$ \gevcc. Figure~\ref{fig:fit} shows the 
results of the fit. The expected background is calculated using
\begin{equation}  
  N_{\rm bkg} = N_{\rm fit} \times \frac{\int^{5.29}_{5.27}A(\mes)d\mes}{\int^{5.24}_{5.10}A(\mes)d\mes} \equiv N_{\rm fit}\times R_{\rm ARGUS}
  \label{eq:new_extrap}
\end{equation}
where $N_{\rm fit}$ is the number of events contributing to the fit. The result is $N_{\rm bkg} = 5.0^{+1.8}_{-1.4}$
events. The uncertainty is determined by varying $\zeta$ by the $\pm 1\sigma$ uncertainty from the fit.
In the MC simulation (scaled to the on-resonance luminosity), we find $5.7\pm0.5$ background events in the signal box, in agreement with the 
data extrapolation. The simulation indicates that the background is primarily continuum, consisting of 57\% light-quark (\uubar, \ddbar, \ssbar), 
23\% \ccbar, and 20\% \BB events. 

By using the ARGUS function to describe the background \mes distribution, we expect to underestimate the contribution of 
backgrounds that peak within the blinded region of \mes. The simulation indicates that only the relatively small component of 
background due to \BB events exhibits a mildly peaking \mes distribution. When the background extrapolation is applied 
to the simulation, the resulting background estimate is $5.2 \pm 0.5$ events, in agreement with the $5.7$ events actually 
found in the signal box. Although neglecting peaking backgrounds could enhance an apparent signal, here the result 
would be a more conservative upper limit. 

We have evaluated the systematic uncertainty in the signal efficiency which includes the muon 
candidate selection (particle identification and tracking efficiency) as well as the reconstruction efficiency of the companion \B.
The muon identification efficiency has been
studied using a muon control sample taken from \epem\to\epem\mumu events in the data. The identification efficiency is measured 
in the control sample in bins of momentum, polar angle, and charge, and the results are incorporated into the 
nominal MC simulation. Due to changing detector conditions, the muon detection efficiency is not stable in time so the simulated events are 
luminosity-weighted with the correct efficiencies for each run period. Averaged over the momentum and polar angle distributions of muons from \bmunu, 
we estimate that the muon identification efficiency for this data sample is 61\% with a systematic uncertainty of 4.2\%. 
The tracking efficiency of the muon candidate was evaluated from the fraction of tracks reconstructed in the SVT that are 
also found in the DCH. We find that the tracking efficiency is overestimated in the simulation by 0.8\%, which is applied as a correction
to the signal efficiency. The associated systematic error is 2\%.
An additional tracking efficiency systematic error of 1\% is included due to the requirement that the event contain at least four 
charged tracks.  

\begin{figure}
  \includegraphics[bb=20 150 500 650, height=7cm]{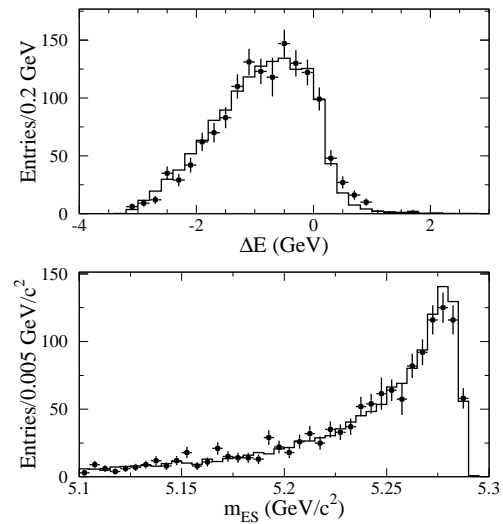}
  \caption{\label{fig:breco} The distributions of \DeltaE and \mes of the companion \B in the $B^+\rightarrow D^0\pi^+$
  control sample after all previous cuts have been applied. The points are the on-resonance data while the histogram is the MC simulation 
  normalized to the number of reconstructed $B^+\rightarrow D^0\pi^+$ decays. 
}
\end{figure}

The companion \B reconstruction efficiency has been studied using a control sample of fully reconstructed
$B^+\rightarrow D^0\pi^+$ and $B^+\rightarrow D^{*0}\pi^+$ events. These are also two-body decays 
in which the \pip momentum spectrum is similar to that of the \mup in signal events. Once reconstructed, 
the pion can be treated as if it were the signal muon and the $D^{(*)0}$ decay products 
can be removed from the event to simulate the unobserved neutrino. Then the companion \B is reconstructed in the control sample as it would be for
signal. We then compare the efficiencies for each of our companion \B selection cuts in the $B^+\rightarrow D^{(*)0}\pi^+$
data and MC simulation. 
Figure~\ref{fig:breco} shows a comparison of on-resonance data and simulation for the \DeltaE and \mes 
distributions in the $B^+\rightarrow D^0\pi^+$ control sample. We expect the resolution observed in the control sample to represent that 
of \bmunu signal events. We find that the efficiency 
after all selection cuts is lower in the data by a factor of 0.94$\pm$0.04 where the uncertainty is due to the statistics 
of the data and MC control samples. Most of this discrepancy is due to the requirement on \mes. The signal efficiency obtained 
from the simulation is therefore corrected by this factor and a systematic error of 4.3\% is applied. A 
summary of the systematic uncertainties in the signal efficiency is given in Table~\ref{tab:systematics}. We estimate the 
overall signal selection efficiency to be 2.09 $\pm$ 0.06 \stat $\pm$ 0.13 \syst \%.

\begin{table}[!htb]
 \caption{ \label{tab:systematics}Contributions to the systematic uncertainty on the signal efficiency. }
 \begin{ruledtabular}
 \begin{tabular}{lcc} 
   source                       & correction & uncertainty   \\
\hline
   tracking efficiency          &            &                \\ 
   \,\, muon                    & 0.992        & 2.0\%         \\
   \,\, companion \B            & -            & 1.0\%         \\
   muon identification          & -            & 4.2\%         \\
   companion \B reconstruction  & 0.94         & 4.3\%         \\
\hline
   total                        & 0.932        & 6.4\%         \\
 \end{tabular}
 \end{ruledtabular}
\end{table} 


\begin{figure}
  \includegraphics[bb=20 30 460 460,height=6cm]{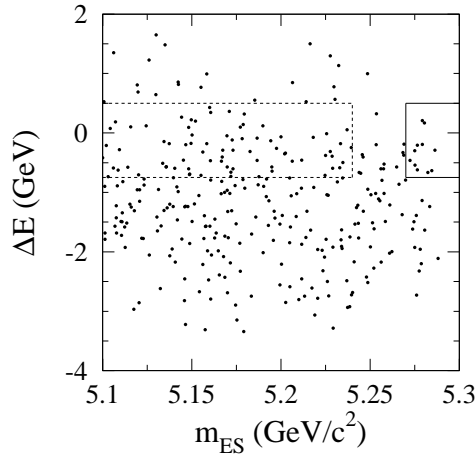}
  \caption{\label{fig:unblind}The distribution of \DeltaE vs \mes in the on-resonance data after 
                              all selection criteria have been applied. The signal box is represented by the solid lines 
                              while the dashed lines indicate the region used to estimate the background.}
\end{figure}

  In the on-resonance data we find 11 events in the signal box where $5.0^{+1.8}_{-1.4}$ background events 
are expected. The distribution of the data in the (\DeltaE, \mes) plane is shown in Figure~\ref{fig:unblind}. 
The 90\% CL upper limit on the number of signal events observed is $n_{UL} = 12.1$ events while the probability of a background 
fluctuation yielding the observed number of events or more is about 4\%.
We set an upper limit on the \bmunu branching fraction using $\BR(\bmunu) < n_{UL}/S$ where $S$ is the sensitivity of the experiment which is the product
of the signal efficiency and the number of \Bpm mesons in the sample. Assuming 
equal production of \Bz and \Bp in \FourS decays, the number of \Bpm mesons in the 
on-resonance data sample is estimated to be 88.4 million with an uncertainty of 1.1\% .   
Systematic uncertainties are included in the upper limit following the prescription given in reference~\cite{Cousins}. We find 
\begin{eqnarray}
 \BR(\bmunu) < 6.6\times10^{-6}\nonumber
\end{eqnarray}
at the 90\% confidence level.


\input pubboard/acknow_PRL.tex

\end{document}

%% file: pubboard/figures.tex
\newif\ifdimspec
\def\figsize#1{\dimspecfalse \checkdim#1\end
\ifdimspec
  \def\figureWidth{#1}%
\else
  \def\figureWidth{#1 in}\fi}
\def\checkdim#1{\ifx#1\end \let\next=\relax
  \else \ifcat#1a \dimspectrue \fi \let\next=\checkdim\fi \next}
\newcommand{\lblcaption  }[2]{\caption{#2\label{\secname#1}}}
\newcommand{\twoAngFiguresEPS}[6]
{
\begin{figure}
\begin{center}
\figsize{#4}
\begin{minipage}[t]{3.2in}
\begin{center}
\epsfig{file=\sectiondir/#1.eps,width=\figureWidth,angle=#5}
\end{center}
\end{minipage}
\begin{minipage}[t]{3.2in}
\begin{center}
\epsfig{file=\sectiondir/#2.eps,width=\figureWidth,angle=#6}
\end{minipage}
\end{center}
\lblcaption{#1}{#3}
\end{figure}
}

%% file: pubboard/authors_oct2003.tex
%
\author{B.~Aubert}
\author{R.~Barate}
\author{D.~Boutigny}
\author{F.~Couderc}
\author{J.-M.~Gaillard}
\author{A.~Hicheur}
\author{Y.~Karyotakis}
\author{J.~P.~Lees}
\author{V.~Tisserand}
\author{A.~Zghiche}
\affiliation{Laboratoire de Physique des Particules, F-74941 Annecy-le-Vieux, France }
% ...
\collaboration{The \babar\ Collaboration}
\noaffiliation

%% file: pubboard/acknow_PRL.tex
We are grateful for the excellent luminosity and machine conditions
provided by our \pep2\ colleagues, 
and for the substantial dedicated effort from
the computing organizations that support \babar.
The collaborating institutions wish to thank 
SLAC for its support and kind hospitality. 
This work is supported by
DOE
and NSF (USA),
NSERC (Canada),
IHEP (China),
CEA and
CNRS-IN2P3
(France),
BMBF and DFG
(Germany),
INFN (Italy),
FOM (The Netherlands),
NFR (Norway),
MIST (Russia), and
PPARC (United Kingdom). 
Individuals have received support from the 
A.~P.~Sloan Foundation, 
Research Corporation,
and Alexander von Humboldt Foundation.